\documentclass[prl,aps,twocolumn]{revtex4-2}
\usepackage[utf8]{inputenc}
\usepackage{graphics, bm, psfrag, amsmath, amssymb, epsfig, grffile, float, bbold}
\usepackage{dsfont}
\usepackage{color}
\usepackage{amsfonts,amsthm,mathtools}
\usepackage{latexsym,graphicx}
\usepackage{amscd}
\usepackage{natbib} 
\usepackage{xcolor}

\newcommand{\intp}{\int\!\! \dd\vk\, }
\newcommand{\intq}{\int\!\! \dd\vk'\, }

\newcommand{\intpq}{\intp\intq}

\renewcommand{\Theta}{\theta}
\newcommand{\ii}{\mathrm{i}}

\newcommand{\dd}{{\rm d}}

\newcommand{\eps}{\epsilon}

\newcommand{\vk}{\mathbf{k}}

\theoremstyle{definition}

\theoremstyle{remark}

\newcommand{\om}{\omega}

\newcommand{\VC}{V_{\mathrm{C}}}
\newcommand{\Vep}{V_{\mathrm{ph}}}
\newcommand{\Vsf}{V_{\mathrm{sf}}}

\begin{document}

\title{Obstructions to odd-frequency superconductivity in Eliashberg theory}

\author{Edwin Langmann}
\affiliation{Department of Physics, KTH Royal Institute of Technology, SE-106 91 Stockholm, Sweden}
\affiliation{Nordita, KTH Royal Institute of Technology and Stockholm University, SE-106 91 Stockholm, Sweden}

\author{Christian Hainzl}
\affiliation{Mathematisches Institut, Ludwig-Maximilians-Universit\"at M\"unchen, 80333 Munich, Germany}

\author{Robert Seiringer}
\affiliation{IST Austria, Am Campus 1, 3400 Klosterneuburg, Austria}

\author{Alexander V. Balatsky}
\affiliation{Nordita, KTH Royal Institute of Technology and Stockholm University, SE-106 91 Stockholm, Sweden}
\affiliation{Department of Physics, University of Connecticut, Storrs, CT 06268, USA}

\date{July 4, 2022} 

\begin{abstract}
We present a necessary condition for odd-frequency (odd-f) superconductivity (SC) to occur in a large class of materials described by Eliashberg theory.
We use this condition to prove a no-go theorem ruling out the occurrence of odd-f SC in standard one-band superconductors with pairing interactions mediated by phonon exchange. 
We also present a corresponding no-go theorem for superconductors with interactions mediated by spin-fluctuations. 
Our results explain why odd-f SC is rare in conventional materials, and they open up the possibility for a search for materials with interactions designed so as to allow for odd-f SC.
\end{abstract}
\maketitle

\noindent {\bf Introduction.} Superconductivity (SC) is a fascinating phenomenon where quantum coherence is developed  on a macroscopic scale. 
The usual narrative is that the condensation of Cooper pairs leads to  coherence that is responsible for the striking phenomena exhibited by superconductors, including resistance-less flow of an electron fluid,  Meissner effect, superconducting vortices, and the Josephson effect.
Ginzburg-Landau theory for SC uses the pair condensate wave function, $\Psi({\bf r},t) = \langle c_\uparrow({\bf r},t)c_\downarrow({\bf r},t)\rangle$, as an order parameter  \cite{leggett2006,svistunov2015} ($c_\sigma({\bf r},t)$ are electron operators at position ${\bf r}$, time $t$, and with spin index $\sigma=\uparrow,\downarrow$). 
This BCS-Ginzburg-Landau paradigm has been tremendously successful; ultimately, it enabled the viable platform for quantum computing based on Josephson junctions \cite{arute2019}. 

With all the success of the current paradigm, there is a major omission in the discussion  of SC: 
it omits the challenge posed by the existence of a whole class called odd-frequency (odd-f) superconductors, see e.g.\ \cite{tanaka2011,linder2019}. 
The name {\em odd-f} is a consequence of highly nontrivial  pairing correlations where  amplitudes 
$F_{\sigma,\sigma'}({\bf r},t|{\bf r'},t')  = \langle c_\sigma({\bf r},t)c_{\sigma'}({\bf r'},t')\rangle$ 
are {\it odd} functions under permutation of (Matsubara imaginary) times $t$ and $t'$;
hence, the equal time order parameter $\Psi({\bf r},t)$, which is proportional to the equal time pairing amplitude, vanishes. 
This type of pairing correlations, along with the title {\it odd-f} (sometimes called Berezinskii pairing), 
 was initially proposed for spin triplet superfluid $^3\mathrm{He}$ by Berezinskii \cite{berezinskii1974}. 
One can extend these pairing states to the case of superconductors with electronic condensate \cite{balatsky1992}.   
By now, we know that odd-f pairing correlations can be induced in SCFM (ferromagnetic heterostructures), in the vortex cores, in superfluid $^3$He near boundaries, in Josephson junctions, in interacting Majorana fermions, and in Dirac semimetals,  provided that the pairing interactions are strong enough \cite{linder2019}. 

With the literature on emergent odd-f pairing states growing, it is still viewed as an exotic and remote possibility. 
Indeed, in a majority of superconducting states, the conventional even-frequency (even-f) BCS pairing channel is dominant.   
Moreover, after unsuccessful attempts to find odd-f SC by solving the Eliashberg equations numerically, it was argued already early on that pairing by phonons is
unlikely to give odd-f SC \cite{abrahams1993}; see also \cite{pimenov2022} for a recent such argument.
Hence the question is: why are odd-f pairing states so rare and hard to stabilize?  
Or more precisely: if the odd-f pairing is a fundamentally allowed pairing state, why is it so hard to observe in nature?
One possible answer is that we simply do not understand  all the  physical properties of odd-f superconductors and how they are different from even-f BCS-like superconductors.
Here, we present an alternative explanation for the relative  scarcity of odd-f pairing states found to date: conventional materials in nature typically have interaction potentials that do not allow for odd-f superconducting solutions of the pertinent Eliashberg equations. 
Since odd-f pairing requires strong interactions, one needs to keep track of both normal selfenergy renormalizations as well as pairing self energies on equal footing. 
We find an obstruction of the gap equation and selfenergy correction to have a non-trivial solution in the odd-f channel for the specific cases of spin independent electron-phonon interaction: 
The strong coupling required to produce odd-f pairing states at the same time leads to a strong selfenergy renormalization that ultimately prohibits odd-f states. 
 We formulate this obstruction as a {\it no-go theorem} for odd-f pairing states within the Eliashberg formalism for specific cases.
This theorem explains  why numerous attempts to produce odd-f states in the Eliashberg framework have failed.
It also clarifies how one would have to change the interactions to allow for odd-f SC.  
We suggest to turn this theorem into a design principle for pairing interactions that lead to odd-f SC states in the Eliashberg approach.  
Thus, the challenge is shifted to finding materials where such an interaction is realized.
 
 It is important to note that our results hold true for the general Eliashberg equations, taking into account the (Matsubara) frequency dependence and momentum dependence of the interactions, without any further approximation.  Commonly, these general Eliashberg equations are simplified by the local approximation, eliminating the momentum dependence \cite{allen1983}; in particular, the original works on odd-f SC were based on these simplified Eliashberg equations \cite{balatsky1992,abrahams1993}. We derive our no-go theorem for the simplified Eliashberg equations in the main text; the extension to the general Eliashberg equations is explained in the main text, with the detailed proof deferred to Appendix~A. For simplicity, we only consider the {\em linearized} Eliashberg equations determining the superconducting critical temperature, $T_c$, in the main text; as shown in Appendix~A, it is easy to drop this restriction. In the main text, we also present another no-go theorem derived in Appendix~A, which applies to systems that, in addition to Coulomb- and phonon interactions, also have interactions mediated by spin fluctuations. 

\medskip 

\noindent {\bf {\em SP$^*$T$^*$}-rule.} 
We start by discussing symmetries that allow to classify different possibilities for superconducting states in one-band models, including odd-f SC \cite{linder2019}. 
We consider models with fermion operators, $c_{\sigma}(\ii\om_n,\vk)$ \cite{comment},  labeled by a spin index, $\sigma=\uparrow,\downarrow$, Masubara frequencies, $\om_n=(2n+1)\pi T$ with integers $n$ and $T>0$ the temperature, and pseudo-momenta in the Brillouin zone, $\vk$. A possible superconducting state is described by the fermion correlation functions
\begin{equation}
\label{F}
F_{\sigma,\sigma'}(\ii\om_n,\vk)\equiv \langle c_{\sigma}(\ii\om_n,\vk)c_{\sigma'}(-\ii\om_n,-\vk)\rangle,
\end{equation}
and $G_{\sigma,\sigma'}(\ii\om_n,\vk)\equiv \langle c^\dag_{\sigma}(\ii\om_n,\vk)c_{\sigma'}(\ii\om_n,\vk)\rangle$, 
assuming thermal equilibrium and translational invariance.
We consider transformations $S$, $P^*$, and $T^*$ that swap the spin index, the sign of the momentum, and the sign of the Matsubara frequency, respectively:
\begin{equation}
\label{SPT} 
\begin{split}
S: \; \sigma\leftrightarrow \sigma', \quad
P^*: \; \vk\to -\vk,\quad
T^*: \; \om_n\to -\om_n.
\end{split}
\end{equation}
Note that all these transformations square to the identity $1$, i.e., the possible eigenvalues of $S$, $P^*$ and $T^*$ are $+1$ and $-1$.
Obviously the combined transformation $SP^*T^*$ interchanges the two fermion operators in \eqref{F} and, by the anti-commutativity of fermion operators, $SP^*T^*= -1$. We are mainly interested in odd-f SC where $T^*=-1$, i.e., $F$ is a odd function of Matsubara frequencies. The one-band models we consider are invariant under the transformation $S$, $P^*$ and $T^*$ separately, which implies that the superconducting gap, $\Delta$, has the same transformation properties under these transformations as $F$. In the odd-f SC case there are two possibilities: $(S,P^*,T^*)=(-,-,-)$ and $(+,+,-)$,  which correspond to spin-singlet odd-parity odd-f SC and spin-triplet even-parity odd-f SC, respectively. 
 We recall for later reference that, in a rotation invariant system, $P^*=+1$ corresponds to $s$- or $d$-wave SC, whereas $P^*=-1$ corresponds to $p$-wave SC.

\bigskip 

\noindent {\bf No-go theorem in special case.} We consider the simplified Eliashberg equations obtained by projection to the $s$-wave channel and ignoring the $|\vk|-$dependence of the interactions (local approximation); see \cite{allen1983,carbotte1990,marsiglio2008,marsiglio2020}. 

The simplified Eliashberg equations for a one-band superconductor in the local approximation determine the superconducting order parameter (also known as gap), $\Delta(\ii\om_n)$, and the normal-state renormalization, $Z(\ii\om_n)$, as follows (we use units such that $k_B=\hbar=1$)
\begin{equation} 
\begin{split} 
\label{E1} 
Z(\ii\om_n)\Delta(\ii\om_n) &= -T\sum_{m} V(\ii\om_n,\ii\om_m)\frac{\pi}{|\om_m|}\Delta(\ii\om_m), \\
Z(\ii\om_n) &=  1-\frac{\pi}{\om_n} T\sum_{m} V(\ii\om_n,\ii\om_m)\frac{\om_m}{|\om_m|}.
\end{split} 
\end{equation}
The standard potential describing interactions mediated by phonons is given by 
\begin{multline} 
\label{V0} 
V(\ii\om_n,\ii\om_m) = \mu^*\theta(1-|\om_n|/\om_c) \theta(1-|\om_m|/\om_c) \\
  - \int_0^\infty \alpha^2F(\Omega) \frac{2\Omega}{\Omega^2+(\omega_n-\omega_m)^2}d\Omega
\end{multline} 
where $\mu^*>0$ is the Coulomb pseudo-potential,  $\theta(x)$ the Heaviside function, $\om_c>0$ an energy cutoff, and $\alpha^2F(\Omega)\geq 0$ the Eliashberg function \cite{allen1983}. 

By writing the gap as a sum of even-f and odd-f parts, $\Delta^+$ and $\Delta^-$, respectively: $\Delta=\Delta^++\Delta^-$ with 
\begin{equation} 
\Delta^\pm(\ii\om_n)\equiv \frac12 [\Delta(\ii\om_n)\pm \Delta(-\ii\om_n)]=\pm\Delta^\pm(-\ii\om_n),
\end{equation} 
the first equation in \eqref{E1} decouples: $Z(\ii\om_n)\Delta^\pm (\ii\om_n) = -T\sum_{m\geq 0} V^\pm (\ii\om_n,\ii\om_m)\pi\Delta^\pm (\ii\om_m)/\om_m$
for $n\geq 0$ with 
\begin{equation} 
\label{Vpm1}
V^\pm(\ii\om_n,\ii\om_m) \equiv V(\ii\om_n,\ii\om_m) \pm V(\ii\om_n,-\ii\om_m)
\end{equation} 
(we used that $V(\ii\om_n,-\ii\om_m)=V(-\ii\om_n,\ii\om_m)$,  which implies that $Z$ is even: $Z(-\ii\om_n)=Z(\ii\om_n)$).
Moreover, the second equation in \eqref{E1} can be written as $Z(\ii\om_n) =  1-(\pi T/\om_n)\sum_{m\geq 0} V^-(\ii\om_n,\ii\om_m)$, 
and by inserting this into the equation for $\Delta^-$, we obtain  
\begin{widetext} 
\begin{equation} 
\label{Tc}
\Delta^-(\ii\om_n) = \pi T\sum_{m\geq 0} V^- (\ii\om_n,\ii\om_m)\left( \frac{\Delta^- (\ii\om_n) }{\om_n} -  \frac{\Delta^- (\ii\om_m) }{\om_m}\right) , 
\end{equation} 
which is an eigenvalue equation of a temperature dependent matrix determining $T_c$.
Multiplying both sides of this equation by the complex conjugate of $\Delta^-(\ii\om_n)/\om_n$, summing over $n\geq 0$, and symmetrizing the resulting double sum on the right-hand side using $V^-(\ii\om_n,\ii\om_m)=V^-(\ii\om_m,\ii\om_n)$ we obtain (further details can be found in Appendix~A.1) 
\begin{equation} 
\label{criterion1} 
T\sum_{n\geq 0} \frac{|\Delta^-(\ii\om_n)|^2}{\om_n} =T^2 \frac{\pi}{2} 
\sum_{m,n\geq 0} V^- (\ii\om_n,\ii\om_m)\left| \frac{\Delta^- (\ii\om_n) }{\om_n} -  \frac{\Delta^- (\ii\om_m) }{\om_m}\right|^2. 
\end{equation} 
Note that this is an {\em exact} consequence of the Eliashberg equations \eqref{E1}. Our  {\bf odd-f-condition} is as follows: {\em To obtain odd-f SC, the interaction potential $V$ must be such that the identity in Eq.\ \eqref{criterion1} does not lead to a contradiction.} 
As we now show, this condition is powerful: 
We compute $V^-$ for the interaction \eqref{V0} and obtain  
\begin{equation} 
V^-(\ii\om_n,\ii\om_m) =
  - \int_0^\infty \alpha^2F(\Omega) \frac{8\Omega\om_n\om_m}{[\Omega^2+(\omega_n-\omega_m)^2][\Omega^2+(\omega_n+\omega_m)^2]}d\Omega \quad (n,m>0), 
\end{equation} 
which is manifestly non-positive (note that minus sign and recall that $\alpha^2F(\Omega)\geq 0$). Clearly, this non-positiveness  leads to a contradiction in \eqref{criterion1} unless $\Delta^-=0$. 
This rules out spin triplet $s$-wave odd-f SC for the simplified Eliashberg equations \eqref{E1}--\eqref{V0}. As explained below, one can adapt this argument to general Eliashberg theory and, by that, rule out odd-f SC  in any channel: our general result is as follows.

\medskip

\noindent {\bf No-go Theorem I:}  {\em In $P^*T^*$-invariant Eliashberg theory, odd-f SC is impossible in a standard one-band superconductor with non-dynamically screened Coulomb interactions and attractive interactions mediated by phonons.} 

\medskip

\noindent {\bf No-go theorem in general case.}  We proceed by explaining the precise meaning of No-go Theorem~I for Eliashberg theory without local approximation. 

The general (linearized) Eliashberg equations which determine the superconducting order parameter, $\phi$, and the normal-state self-energy, $\Sigma$, depending on (Matsubara) frequency, $\om_n$, and momentum, $\vk$, are given by 
\begin{equation}
\label{phiSigma}
\begin{split}
\phi (\ii\om_n,\vk) &= T\sum_m \intq V (\ii\om_n ,\ii\om_m;\vk,\vk')F (\ii\om_m,\vk'),\\
\Sigma (\ii\om_n,\vk) &= -T\sum_m \intq V (\ii\om_n ,\ii\om_m;\vk,\vk')G (\ii\om_m,\vk'),
\end{split}
\end{equation}
with $\intq$ short for $\int\frac{d^3k'}{(2\pi)^3}$, where the superconducting and normal state parts of the fermion correlations functions, $F$ and $G$, are determined as follows,  
$F(\ii\om_n,\vk) = -\phi(\ii\om_n,\vk)/[\tilde\om_n(\vk)^2+\tilde\eps(\ii\om_n,\vk)^2]$ and 
$G(\ii\om_n,\vk) = 1/[\ii \tilde\om_n(\vk)-\tilde\eps(\ii\om_n,\vk)]$; 
see \cite{allen1983,carbotte1990,marsiglio2008} and references therein. 
We use standard notation: $\tilde\om_n(\vk) \equiv \om_nZ(\ii\om_n ,\vk)$ and $\tilde\eps(\ii\om_n,\vk) \equiv \eps(\vk)+\chi (\ii\om_n,\vk)$
with the mass and band renormalizations, $Z$ and $\chi$,  determined by the self-energy as follows \cite{allen1983}, $1-Z(\ii\om_n,\vk) \equiv [\Sigma(\ii\om_n,\vk) -\Sigma(-\ii\om_n,\vk)]/2\ii\om_n$ and 
$\chi (\ii\om_n,\vk) \equiv [\Sigma(\ii\om_n,\vk) +\Sigma(-\ii\om_n,\vk) ]/2$. 
Note that the model is defined by the band relation, $\eps(\vk)$, and the two-body potential, $V=V(\ii\om_n,\ii\om_m;\vk,\vk')$; see Appendix~A.2 for examples. 
Moreover, the equations hold true for spin-singlet and spin-triplet SC, and for $s$-, $p$-, $d$- or $f$-wave SC \cite{allen1983}. 
The only assumption we make is invariance under the transformations $T^*$ and $P^*$, i.e.,
\begin{equation}
\label{Vassumption}
V(\ii\om_n,\ii\om_m;\vk,\vk')   =V(-\ii\om_n,-\ii\om_m;\vk,\vk')  = V(\ii\om_n,\ii\om_m;-\vk,-\vk')
\end{equation} 
and $\eps(\vk)=\eps(-\vk)$. 
We also recall the definition of the superconducting gap  \cite{allen1983}, 
 $\Delta(\ii\om_n,\vk)\equiv \phi(\ii\om_n,\vk)/Z(\ii\om_n,\vk)$.

Similarly as in the simplified case with the local approximation above, one can show that these Eliashberg equations lead to decoupled equations for the even-f and odd-f gap solutions $\Delta^+$ and $\Delta^-$, respectively, with $\Delta^-$ satisfying the following exact identity 
\label{TV}
\begin{equation}
\label{criterion2}
T\sum_{n\geq 0}\intp \frac{|\Delta ^-(\ii\om_n,\vk)|^2}{W(\ii\om_n,\vk)\om_n} = T^2\sum_{n,m\geq 0}\intpq
\frac{V ^-(\ii\om_n,\ii\om_m;\vk,\vk')}{2W(\ii\om_n,\vk)W(\ii\om_m,\vk')}
\Bigl| \frac{\Delta ^-(\ii\om_n,\vk)}{\om_n}  - \frac{\Delta ^-(\ii\om_m,\vk')}{\om_m} \Bigr|^2, 
\end{equation}
$V^-(\ii\om_n,\ii\om_m;\vk,\vk')\equiv V(\ii\om_n,\ii\om_m;\vk,\vk') - V(-\ii\om_n,\ii\om_m,\vk,\vk')$, $W(\ii\om_m,\vk')\equiv \left[ \tilde\om(\vk')^2+\tilde\eps(\ii\om_m,\vk')^2\right]/\om_m Z(\ii\om_m,\vk')$; the interested reader can find a detailed derivation of this fact in Appendix~A.1; see Proposition~1. 
\end{widetext} 
Thus, we get, again, the odd-f condition as formulated above with Eq.~\eqref{criterion1} replaced by Eq.~\eqref{criterion2}. Moreover, similarly as for the special case discussed in more detail above, for standard potentials, $V$,  which describe screened Coulomb interactions and attractive interactions mediated by phonons, one finds that $V^-$ is manifestly non-positive, and this proves No-go theorem~I as it stands for the general Eliashberg equations; the interested reader can find details of our proof in Appendix~A.2. We stress that this no-go theorem is very general: it is true for any kind of odd-f SC, i.e., the spin-structure and angular dependence of the gap make no difference. 

\medskip 

\noindent {\bf Spin dependent interactions.}  Our results above generalize to models with spin dependent interaction potentials. In Appendix~A.1, we discuss more general model potentials which not only include screened Coulomb interactions, $\VC$,  and phonon interactions, $\Vep$,  but also interactions mediated by spin fluctuations, $\Vsf$. In this case, we get another no-go theorem (No-go Theorem~II in Appendix~A.3) ruling out spin-triplet odd-f SC in a one-band superconductor with phonon- and spin-fluctuation mediated interactions. 

Recall that, for odd-f SC,  spin-singlet corresponds to $S=-1$ and $P^*=-1$, and spin-triplet corresponds to $S=+1$ and $P^*=+1$. 
Thus, the no-go theorem allows for spin-singlet $p$-wave odd-f SC, in agreement with results in \cite{fuseya2003}. 
However, by our result, spin-triplet  $s$- and $d$- wave odd-f SC are impossible. 

\medskip

\noindent {\bf Interactions leading to odd-f SC.}  Our results above rule out odd-f SC in a large class of models describing real materials. However, they  can also be used to design interactions leading to odd-f SC. To give examples, we mainly restrict ourselves to the Eliashberg equations \eqref{E1} for simplicity (local approximation and spin-independent interactions). 

A large class of examples avoiding our no-go theorems are repulsive interactions, $V(\ii\om_n,\ii\om_m)\geq 0$,  which are monotonically decreasing functions of $|\om_n-\om_m|$. 
We study three such examples. Our first example is $V(\ii\om_n,\ii\om_m)  = \lambda \Omega^2/[\Omega^2+(\om_n-\om_m)^2]$ for $\lambda>0$ ($\Omega>0$ is an energy scale, here and in the following); this is like a phonon interaction, except that the coupling constant has the opposite sign. To obtain a more physical such example, we start with the momentum-dependent potential given in  \cite[Eq.~(2)]{wu2022} describing screened Coulomb interactions close to a quantum critical point, 
\begin{equation} 
\label{VQCP} 
V(\ii\om_n,\ii\om_m;\vk,\vk')=\frac{g_{\text{eff}}}{|\vk-\vk'|^2+\Gamma\frac{|\om_n-\om_m|}{|\vk-\vk'|}} 
\end{equation} 
for $g_{\text{eff}}>0$ and $\Gamma>0$;  see \cite{brando2016} for other systems described by this and similar such potentials. We note that this potential satisfies our odd-f condition after \eqref{criterion2}. 
In three dimensions, by restricting the momenta to the Fermi surface, $|\vk|=|\vk'|=k_F$ (Fermi momentum),  
and angular averaging, one can reduce this potential to $V(\ii\om_n,\ii\om_m) = \lambda\log(1+\Omega/|\om_n-\om_m|)$ with $\lambda=g_{\text{eff}}N(0)/6k_F^2>0$  and $\Omega=8k_F^3/\Gamma$, where $N(0)$ is the electronic density of states at the Fermi surface (this standard approximation reducing the general Eliashberg equation to \eqref{E1} is discussed in \cite{allen1983}); the latter potential is our second example. Our third example is the potential $V(\ii\om_n,\ii\om_m)=\lambda(\Omega/|\om_n-\om_m|)^\gamma$ for $\gamma>0$, which is similar to the potential of the $\gamma$-model studied recently in \cite{abanov2020,wu2022} but with the opposite sign of the coupling constant (we rename $g_{\text{eff}}$ to $\Omega$ and introduce a coupling constant $\lambda>0$). We solved the $T_c$-equation \eqref{Tc} for these three examples numerically and found robust odd-f solutions, with $T_c$ increasing monotonically with the coupling constant in all three examples. We also obtained a simple exact $T_c$-equation for the third example: $T_c=K\Omega \lambda^{1/\gamma}$, where $K$ is a constant that can be computed numerically. The interested reader can find details and numerical results in Appendix~B. 

\medskip 

\noindent {\bf Conclusions.} 
We present a new constraint on pairing interactions to allow for odd-f solutions within the Eliashberg framework (odd-f condition).
Odd-f SC requires strong coupling to stabilize the odd-f channel and favor it over even-f channel. 
However, both anomalous and {\it normal} selfenergies are renormalized by interactions. 
By combining the corresponding two Eliashberg equations,  we find a strong constraint on pairing interactions: for spin independent interactions mediated by phonons, we find  odd-f solutions are impossible within the Eliashberg formalism (No-go theorem I). 
We also consider spin dependent interactions mediated by spin-fluctuations, and we find that single band models cannot support $s$-wave odd-f solutions. 
Since our no-go theorems apply to a large class of models used to describe real materials, they explain why it has been so difficult to find odd-f SC. 
 
We note that multiband systems do not suffer from this constraint and hence represent a fertile ground to pursue odd-f solutions; for experimentally observable signatures of odd-frequency pairing in multiband superconductors, see e.g. \cite{komendova2015,asano2015,komendova2017,sukhachov2019,triola2020} and references therein.

Our no-go theorems are about models with standard interactions occurring in nature, but there is no general mathematical obstruction to odd-f SC if one drops this latter restriction. 
As we show in examples, using our results, one can construct models with pairing interactions that allow for odd-f SC. 
Thus, our results can be used as a design principle to construct models for odd-f SC which, as we hope, can guide the search for real odd-f materials. 
 
Our results are an example of how mathematics can contribute to the theory of superconductivity; see, e.g., \cite{hainzl2008,hainzl2016,triola2019} for recent other such examples. 
 
{\it Acknowledgements.} We are grateful to A. Black-Schaffer, M. Geilhufe, P. Coleman, P. Sukhachov, and C. Triola,  for useful discussions. 
We thank A. Chubukov for pointing out an important detail in Ref.\ \cite{wu2022} and valuable comments. 
The work of AVB was  supported by VILLUM FONDEN via the Centre of Excellence for Dirac Materials (Grant No.  11744) and the European Research Council ERC-2018-SyG HERO.

\newpage 

\appendix
 
\begin{widetext}

\section{Appendix A. Proof of no-go theorem for general Eliashberg equations} 
We give detailed derivations of our results for the general case: a $P^*T^*$-invariant one-band model given by a dispersion relation, $\eps$,  and two-body interaction potentials,  $V$ and $\tilde V$. As explained in Appendix~A.2, for conventional superconductors with Coulomb interactions and attractive interactions due to phonons, $V$ is the same for spin singlet and spin triplet SC and $V=\tilde{V}$; see \eqref{Vconv}. However, our derivation also applies to non-conventional superconductors discussed in Appendix~A.3 where, in addition to Coulomb- and phonon interactions, also attractive interactions due to spin fluctuations are present. In the latter case, there are different potentials $V$ for the spin singlet and spin triplet cases, and $\tilde{V}$ differs from $V$; see \eqref{Vnonconv1}--\eqref{Vnonconv2}.  One specific example for $\eps$ would be the Sommerfeld dispersion relation,  $\eps(\vk)=\vk^2/2m^*-\mu$ with the effective mass $m^*$ and the chemical potential $\mu$. However, we stress that our results hold true also also for non-rotation invariant systems and systems with a finite Brillouin zone. 

\subsection{A.1 Derivation of odd-f condition}
In the general case, the Eliashberg equations given in the main text have to be generalized by replacing \eqref{phiSigma} with  
\begin{equation}
\label{phiSigmagen} 
\begin{split}
\phi (\ii\om_n,\vk) &= T\sum_m \intq V (\ii\om_n ,\ii\om_m;\vk,\vk')F (\ii\om_m,\vk'),\\
\Sigma (\ii\om_n,\vk) &= -T\sum_m \intq \tilde{V} (\ii\om_n ,\ii\om_m;\vk,\vk')G (\ii\om_m,\vk')
\end{split}
\end{equation}
(our notation is explained in the main text), while we use generalizations of formulas for $F$ and $G$ in the main text to $0\leq T\leq T_c$, 
\begin{equation}
\label{FGgen}
\begin{split}
F(\ii\om_n,\vk) = -\frac{ \phi(\ii\om_n,\vk)}{\tilde\om_n(\vk)^2+\tilde\eps(\ii\om_n,\vk)^2+|\phi(\ii\om_n,\vk)|^2},\\
G(\ii\om_n,\vk) = -\frac{ \ii \tilde\om_n(\vk)+\tilde\eps(\ii\om_n,\vk)}{\tilde\om_n(\vk)^2+\tilde\eps(\ii\om_n,\vk)^2+|\phi(\ii\om_n,\vk)|^2}
\end{split}
\end{equation}
(note that,  for $|\phi(\ii\om_n,\vk)|^2=0$,  this simplifies to the formulas for $F$ and $G$ given in the main text); the definitions 
\begin{equation} 
\label{tomnteps}
\tilde\om_n(\vk) \equiv \om_nZ(\ii\om_n ,\vk), \quad \tilde\eps(\ii\om_n,\vk) \equiv \eps(\vk)+\chi (\ii\om_n,\vk),\quad \Delta(\ii\om_n,\vk) = \frac{\phi(\ii\om_n,\vk) }{Z(\ii\om_n,\vk)}, 
\end{equation} 
with 
\begin{equation}
\label{Zchi} 
\begin{split}
1-Z(\ii\om_n,\vk) &\equiv \frac1{2\ii\om_n}[\Sigma(\ii\om_n,\vk) -\Sigma(-\ii\om_n,\vk) ],\\
\chi (\ii\om_n,\vk) &\equiv \frac12 [\Sigma(\ii\om_n,\vk) +\Sigma(-\ii\om_n,\vk) ]
\end{split}
\end{equation}
are as in the main text; see \cite{allen1983,carbotte1990,marsiglio2008,marsiglio2020} and references therein for background to these equations. 
Note that the spin dependence is eliminated from these equations, and they hold true for spin-singlet and spin-triplet SC  with interaction potentials given below. 
We assume invariance under the transformations $P^*$ and $T^*$, i.e., we assume  \eqref{Vassumption} and the same relations with $\tilde{V}$ instead of $V$ hold true, and, in addition, $\eps(-\vk)=\eps(\vk)$. 

We use the definition of the gap $\Delta$ in \eqref{tomnteps} and, similarly as before, we divide the gap in even-f and odd-f parts,
\begin{equation} 
\label{Deltapm} 
\Delta(\ii\om_n,\vk) = \Delta^+(\ii\om_n,\vk) + \Delta^-(\ii\om_n,\vk),\quad \Delta^\pm (\ii\om_n,\vk) = \frac12\left(\Delta(\ii\om_n,\vk)  \pm \Delta(-\ii\om_n,\vk)  \right) = \pm \Delta^\pm (-\ii\om_n,\vk) . 
\end{equation} 
In the following, we use the following definitions of the even- and odd parts of the interaction potential $V$, 
\begin{equation}
\label{Vpm}  
V^\pm(\ii\om_n,\ii\om_m;\vk,\vk')\equiv V(\ii\om_n,\ii\om_m;\vk,\vk') \pm V(-\ii\om_n,\ii\om_m,\vk,\vk'), 
\end{equation} 
and similarly for $\tilde{V}$.

\medskip

\noindent {\bf Lemma~1:} {\em The general Eliashberg equations above imply 
\begin{equation} 
\label{TZchi}
Z(-\ii\om_n,\vk)= Z(\ii\om_n,\vk), \quad \chi(-\ii\om_n,\vk)= \chi(\ii\om_n,\vk), 
\end{equation} 
and 
\begin{equation}
\begin{split}
\label{DeltaZchi}
Z(\ii\om_n,\vk) \Delta ^\pm(\ii\om_n,\vk) & = - T\sum_{m\geq 0}\intq \frac{V ^\pm(\ii\om_n,\ii\om_m;\vk,\vk')}{W(\ii\om_m,\vk')}\frac{\Delta ^\pm(\ii\om_m,\vk')}{\om_m},\\
\chi(\ii\om_n,\vk) & = T\sum_{m\geq 0}\intq \frac{\tilde{V} ^+(\ii\om_n,\ii\om_m;\vk,\vk')}{W(\ii\om_m,\vk')}\frac{\tilde\eps(\ii\om_m,\vk')}{\om_mZ(\ii\om_m,\vk')},\\
Z(\ii\om_n,\vk) & = 1-\frac{T}{\om_n}\sum_{m\geq 0}\intq \frac{\tilde{V} ^-(\ii\om_n,\ii\om_m;\vk,\vk')}{W(\ii\om_m,\vk')}, 
\end{split}
\end{equation}
with 
\begin{equation} 
\label{Wdefgen} 
W(\ii\om_m,\vk')\equiv \frac{\tilde\om(\vk')^2+\tilde\eps(\ii\om_m,\vk')^2+|\phi(\ii\om_n,\vk)|^2 }{\om_m Z(\ii\om_m,\vk') };
\end{equation} 
for $T=T_c$,  these results holds true for any solution of the Eliashberg equations, and for $0<T<T_c$ they holds true for any solution such that either $\Delta=\Delta^+$ or $\Delta=\Delta^-$. 
} 
\medskip 

Note that, by the transformation properties of $\Delta^\pm$, $Z$ and $\chi$ under $T^*:\; \om_n\to -\om_n$, one can restrict \eqref{DeltaZchi} to positive Matsubara frequences, $n\geq 0$, and $W(\ii\om_m,\vk')\geq 0$ for all $m\geq 0$.  

\medskip

\noindent {\bf Remark~1:} To explain the distinctions between $T=T_c$ and $T<T_c$ in Lemma~1, we note that the linearized Eliashberg equations determining $T_c$ are obtained from this by simplifying $W(\ii\om_m,\vk')$ in \eqref{Wdefgen} to 
\begin{equation} 
\label{Wdef} 
W(\ii\om_m,\vk')\equiv \frac{\tilde\om(\vk')^2+\tilde\eps(\ii\om_m,\vk')^2}{\om_m Z(\ii\om_m,\vk') }, 
\end{equation} 
and this satisfies $W(\ii\om_m,\vk')=-W(-\ii\om_m,\vk')$ without further assumption; however, for $T<T_c$, this only holds true if  $|\phi(\ii\om_n,\vk)|^2 = |Z(\ii\om_n,\vk)\Delta(\ii\om_n,\vk)|^2 = |\phi(-\ii\om_n,\vk)|^2$ and, for this reason, our results hold  true only if $\Delta=\Delta^\pm$. Thus, in principle, our no-go theorems do not rule out the possibility of a mixed gap at temperatures below $T_c$, i.e., they leave open the possibility that, at $T<T_c$, $\Delta=\Delta^++\Delta^-$ with $\Delta^+$ and $\Delta^-$ both non-zero.

 \medskip

\begin{proof}[Proof of Lemma~1]
In the following, we use the notation $\phi^\pm(\ii\om_n,\vk)\equiv [\phi(\ii\om_n,\vk)\pm \phi(-\ii\om_n,\vk)]/2$ and similarly for $\Sigma$, $F$ and $G$.  

The symmetry properties stated in \eqref{TZchi} are consequences of  the definitions of $Z$ and $\chi$ in \eqref{Zchi}. 
Moreover,  \eqref{FGgen}, \eqref{tomnteps} and  \eqref{TZchi}  imply $\phi^\pm(\ii\om_n,\vk)=Z(\ii\om_n,\vk)\Delta^\pm(\ii\om_n,\vk)$ and 
\begin{equation} 
\label{FpmGpm} 
F^\pm(\ii\om_n,\vk) = -\frac{\Delta^\pm(\ii\om_n,\vk)}{\om_nW(\ii\om_n,\vk)},\quad G^+(\ii\om_n,\vk) = -\frac{\tilde\eps(\ii\om_n,\vk)}{\om_nZ(\ii\om_n,\vk)W(\ii\om_n,\vk)},\quad 
G^-(\ii\om_n,\vk) = -\frac{\ii}{W(\ii\om_n,\vk)}.
\end{equation} 

The first equation in \eqref{phiSigmagen} and the definitions of $\phi^\pm$ and $V^\pm$ imply 
\begin{equation}
\begin{split}  
\phi^\pm(\ii\om_n,\vk) = & T\sum_{m} \intq \frac 12V^\pm(\ii\om_n,\ii\om_m;\vk,\vk')F(\ii\om_m,\vk') \\
& = T\sum_{m\geq 0}\intq \frac12 \left[ V^\pm(\ii\om_n,\ii\om_m;\vk,\vk')F(\ii\om_m,\vk')+V^\pm(\ii\om_n,-\ii\om_m;\vk,\vk')F(-\ii\om_m,\vk') \right]  . 
\end{split} 
\end{equation} 
Inserting $V^\pm(\ii\om_n,-\ii\om_m;\vk,\vk') =\pm V^\pm(\ii\om_n,\ii\om_m;\vk,\vk')$ implied by \eqref{Vassumption} and \eqref{Vpm}, we compute this as follows, 
\begin{equation}
\begin{split}  
\phi^\pm(\ii\om_n,\vk)  & =  T\sum_{m\geq 0}\intq V^\pm(\ii\om_n,\ii\om_m;\vk,\vk')\frac12[F(\ii\om;m,\vk') \pm F(-\ii\om,\vk') ] \\
& =  T\sum_{m\geq 0}\intq V^\pm(\ii\om_n,\ii\om_m;\vk,\vk')F^\pm(\ii\om_m,\vk) =  -T\sum_{m\geq 0} \intq V^\pm(\ii\om_n,\ii\om_m;\vk,\vk')\frac{\Delta^\pm(\ii\om_n,\vk)}{\om_nW(\ii\om_n,\vk)}, 
\end{split} 
\end{equation} 
recalling the definition of $F^\pm$ in the second identity and inserting $F^\pm$ in \eqref{FpmGpm} in the third. 
Since $\phi^\pm(\ii\om_n,\vk)=Z(\ii\om_n,\vk)\Delta^\pm(\ii\om_n,\vk)$, this proves the first equation in \eqref{DeltaZchi}.  

In the same way, the second equation in \eqref{phiSigmagen} implies 
$\Sigma^\pm(\ii\om_n,\vk)= -T\sum_{m\geq 0}\intq  \tilde{V}^\pm(\ii\om_n,\ii\om_m;\vk,\vk')G^\pm(\ii\om_m,\vk')$, 
and thus with \eqref{Zchi} and \eqref{FpmGpm}, 
\begin{equation} 
\begin{split} 
\chi(\ii\om_n,\vk) & = \Sigma^+(\ii\om_n,\vk) = -T\sum_{m\geq 0}\intq \tilde{V}^+(\ii\om_n,\ii\om_m;\vk,\vk')G^+(\ii\om_m,\vk') \\ & = -T\sum_{m\geq 0}\intq \tilde{V}^+(\ii\om_n,\ii\om_m;\vk,\vk')
\frac{\tilde\eps(\ii\om_m,\vk')}{\om_nZ(\ii\om_m,\vk')W(\ii\om_m,\vk')}
\end{split} 
\end{equation} 
and 
\begin{equation}  
\begin{split}  
Z(\ii\om_n,\vk) & =  1 - \frac1{\ii\om_n}\Sigma^-(\ii\om_n,\vk) =1 + \frac{T}{\ii\om_n}\sum_{m\geq 0}\intq \tilde{V}^-(\ii\om_n,\ii\om_m;\vk,\vk')G^-(\ii\om_m,\vk') \\ 
& =  1 - \frac{T}{\om_n}\sum_{m\geq 0}\intq \frac{\tilde{V}^-(\ii\om_n,\ii\om_m;\vk,\vk')}{W(\ii\om_m,\vk')}.
\end{split} 
\end{equation} 
This proves the second and the third equations in \eqref{DeltaZchi}.  
\end{proof} 

We now are ready to state and prove the odd-f condition for the general Eliashberg equations. 

\medskip 

\noindent {\bf Proposition~1} (Odd-f Condition): {\em The equations in \eqref{DeltaZchi} imply 
\begin{equation}
\label{criterion3}
\begin{split} 
T\sum_{n\geq 0}\intp \frac{|\Delta ^-(\ii\om_n,\vk)|^2}{W(\ii\om_n,\vk)\om_n} = T^2\sum_{n,m\geq 0}\intpq\Biggl[
\frac{V ^-(\ii\om_n,\ii\om_m;\vk,\vk')}{2W(\ii\om_n,\vk)W(\ii\om_m,\vk')}
\Bigl| \frac{\Delta ^-(\ii\om_n,\vk)}{\om_n}  - \frac{\Delta ^-(\ii\om_m,\vk')}{\om_m} \Bigr|^2 \\
+ \frac{(\tilde{V}-V) ^-(\ii\om_n,\ii\om_m;\vk,\vk')}{W(\ii\om_n,\vk)W(\ii\om_m,\vk')}
\Bigl| \frac{\Delta ^-(\ii\om_n,\vk)}{\om_n} \Bigr|^2\Biggr]
\end{split} 
\end{equation}
where
\begin{equation} 
(\tilde{V}-V) ^-(\ii\om_n,\ii\om_m;\vk,\vk')  \equiv \tilde{V} ^-(\ii\om_n,\ii\om_m;\vk,\vk') -V ^-(\ii\om_n,\ii\om_m;\vk,\vk'). 
\end{equation} 
Thus, to to obtain odd-f SC, the interactions $V$ and $\tilde{V}$ must be such that \eqref{criterion3} does not lead to a contradiction. 
} 
\medskip 

Note that, for $\tilde{V}=V$ and $W$ in \eqref{Wdef} (which corresponds to spin-independent interactions at $T=T_c$), \eqref{criterion3} reduces to \eqref{criterion2} in the main text. 

\medskip

\begin{proof}[Proof of Proposition~1:] 
In our argument below, we use the following property of interaction potentials $V$: 
\begin{equation} 
\label{symmetry}
V(\ii\om_n,\ii\om_m;\vk,\vk')= V(\ii\om_m,\ii\om_n;\vk',\vk). 
\end{equation} 
It is important to note that this always holds true (for the convenience of the reader, we recall the argument for spin independent interactions: in the functional integral formalism (see e.g.\ Ref.~\onlinecite{salmhofer2007}), the interaction part of the action defining the interacting fermion model is $\frac 12T^2\sum_{n,m}\intpq V(\ii\om_n,\ii\om_m;\vk,\vk') \hat{n}(\ii\om_n,\vk)\hat{n}(\ii\om_m;\vk')$ where $\hat{n}(\ii\om_n,\vk)$ are Fourier transformed fermion densities which commute and thus, clearly, one can assume \eqref{symmetry} without loss of generality). 

We find it convenient to abbreviate the variables $(\om_n,\vk)$ and $(\om_m,\vk')$ by $N$ and $M$, respectively, and $T\sum_{m\geq 0}\intq$ by $\sum_M$ etc.
This allows us to write the first and third equations in \eqref{DeltaZchi} in the odd-f case short as
\begin{equation}
\label{abstract}
Z_N \Delta^-_N  = -\sum_M \frac{V^-_{NM}}{W_M}\frac{\Delta^-_M}{\om_M}, \quad 
Z_N  = 1-\frac1{\om_N} \sum_M \frac{\tilde{V}^-_{NM}}{W_M}
\end{equation}
with $\om_N\equiv \om_n$, $Z_N\equiv Z(\ii\om_n,\vk)$, $V^-_{NM}\equiv V^-(\ii\om_n,\ii\om_m;\vk,\vk')$ etc. 
Inserting the second in the first equation we obtain 
\begin{equation} 
\Delta_N^-  -\sum_M \frac{\tilde{V}^-_{NM}}{W_M} \frac{\Delta_N^- }{\om_N} = -\sum_M \frac{V^-_{NM}}{W_M}\frac{\Delta^-_M}{\om_M}, 
\end{equation} 
which is equivalent to 
\begin{equation}
\Delta_N^-  = \sum_M \left[  \frac{V^-_{NM}}{W_M}\left(\frac{\Delta^-_N}{\om_N} - \frac{\Delta^-_M}{\om_M} \right) + \frac{(\tilde{V}-V)^-_{NM}}{W_M}\frac{\Delta^-_N}{\om_N} \right]
\end{equation}
where $(\tilde{V}-V)^-_{NM} \equiv  \tilde{V}^-_{NM} - V^-_{NM}$. 
Multiplying both sides of this identity with $\overline{\Delta^-_N}/W_N\om_N$ and summing over $N$ we find (the bar indicates complex conjugation), 
\begin{equation}
\label{sum}
\sum_N\frac{|\Delta^-_N|^2}{W_N\om_N} = \sum_{M,N} \left[  \frac{V^-_{NM}}{W_NW_M}\left(\frac{\Delta^-_N}{\om_N} - \frac{\Delta^-_M}{\om_M} \right)\frac{\overline{\Delta^-_N}}{\om_N}
+ \frac{(\tilde{V}-V)^-_{NM}}{W_NW_M}\left|\frac{\Delta^-_N}{\om_N}\right|^2 \right] .
\end{equation}
We observe that \eqref{Vassumption},  the definition of $V^-$ in \eqref{Vpm}, and \eqref{symmetry} imply that $V^-_{NM}$ is even under the exchange $N\leftrightarrow M$: $V^-_{MN}=V^-_{NM}$. 
We use this to anti-symmetrize the first sum on the right-hand side in \eqref{sum}: 
\begin{equation} 
 \sum_{M,N} \frac{V^-_{NM}}{W_NW_M}\left(\frac{\Delta^-_N}{\om_N} - \frac{\Delta^-_M}{\om_M} \right)\frac{\overline{\Delta^-_N}}{\om_N}
 =  \sum_{N,M} \frac{V^-_{MN}}{W_MW_N}\left(\frac{\Delta^-_M}{\om_M} - \frac{\Delta^-_N}{\om_N} \right)\frac{\overline{\Delta^-_M}}{\om_M} = 
  - \sum_{M,N} \frac{V^-_{NM}}{W_NW_M}\left(\frac{\Delta^-_N}{\om_N} - \frac{\Delta^-_M}{\om_M} \right)\frac{\overline{\Delta^-_M}}{\om_M} 
\end{equation} 
(in the first identity, we only renamed summation variables $(N,M)\to (M,N)$), and by equating the left-hand side with the average of the left- and right-hand sides of this identity we obtain 
\begin{equation} 
\begin{split} 
 \sum_{M,N} \frac{V^-_{NM}}{W_NW_M}\left(\frac{\Delta^-_N}{\om_N} - \frac{\Delta^-_M}{\om_M} \right)\frac{\overline{\Delta^-_N}}{\om_N}= & 
  \frac12\sum_{M,N} \frac{V^-_{NM}}{W_NW_M}\left(\frac{\Delta^-_N}{\om_N} - \frac{\Delta^-_M}{\om_M} \right)\left(\frac{\overline{\Delta^-_N}}{\om_N}- \frac{\overline{\Delta^-_M}}{\om_M} \right) \\
  =  & \sum_{M,N} \frac{V^-_{NM}}{2W_NW_M}\left|\frac{\Delta^-_N}{\om_N} - \frac{\Delta^-_M}{\om_M} \right|^2. 
\end{split} 
\end{equation} 
Inserting this into \eqref{sum}, we obtain 
\begin{equation}
\label{result}
\sum_N \frac{|\Delta^-_N|^2}{W_N\om_N} = \sum_{M,N}\Biggl[  \frac{V^-_{NM}}{2W_NW_M}\left|\frac{\Delta^-_N}{\om_N} - \frac{\Delta^-_M}{\om_M} \right|^2 
+  \frac{(\tilde{V}-V)^-_{NM}}{W_NW_M}\left|\frac{\Delta^-_N|}{\om_N}\right|^2\Biggr] ,
\end{equation}
which is the identity in \eqref{criterion3}.

We showed that  \eqref{criterion3} is an exact consequence of the Eliashberg equations. Thus, clearly, any odd-f solution $\Delta=\Delta^-$ of the Eliashberg equations must satisfy \eqref{criterion3}. 
\end{proof} 

\subsection{A.1 Spin independent interactions} 
We first consider a standard superconductor where the electrons interact with screened Coulomb interactions, $\VC(\vk,\vk')$,  independent of the frequencies $\om_n$ and $\omega_m$, and with attractive interactions mediated by phonons,
\begin{equation}
\label{Vep}
\Vep(\ii\om_n,\ii\om_m;\vk,\vk')=- \frac1{N(0)} \int_0^\infty \alpha^2F(\Omega;\vk,\vk') \frac{2\Omega}{\Omega^2+(\om_n-\om_m)^2} d\Omega,
\end{equation}
with $N(0)$ the electron density at the Fermi surface and $\alpha^2F(\Omega;\vk,\vk')$ the Eliashberg function; see \cite{allen1983}.
Ignoring impurities one has, in this case, 
\begin{equation} 
\label{Vconv} 
V=\tilde{V}=\VC+\Vep
\end{equation} 
(note that, since $(\tilde{V}-V)^-=0$, \eqref{criterion3} simplifies to  \eqref{criterion2}). By simple computations we find $(\VC) ^-=0$ (since $\VC$ is frequency independent) and  
\begin{equation}
\label{Vep1}
(\Vep) ^-(\ii\om_n,\ii\om_m;\vk,\vk') =  - \frac1{N(0)}\int_0^\infty \alpha^2F(\Omega;\vk,\vk')\frac{8\Omega\om_n\om_m}{[\Omega^2+(\om_n-\om_m)^2][\Omega^2+(\om_n+\om_m)^2]}d\Omega \quad (n,m\geq 0).
\end{equation}
We recall that the Eliashberg function is determined by the phonon frequencies, $\om_\lambda(\vk)$, and electron-phonon coupling constants, $g_\lambda(\vk,\vk')$, as follows, $\alpha^2F(\Omega;\vk,\vk') = \sum_{\lambda} |g_\lambda(\vk,\vk')|^2\delta(\Omega-\om_\lambda(\vk-\vk'))$,  and thus, $\alpha^2F(\Omega;\vk,\vk')\geq 0$. This implies that $(\Vep) ^-(\ii\om_n,\ii\om_m;\vk,\vk')$ in \eqref{Vep1} is non-positive. 
Since $(\tilde{V}-V) ^-=0$, $V ^-=(\Vep) ^-\leq 0$,  and $W(\ii\om_m,\vk')\geq 0$ for $n,m\geq 0$, the RHS in \eqref{criterion3} is non-positive, which is in contradiction with the LHS in  \eqref{criterion3} which is manifestly non-negative. The only way to avoid this contradiction is $\Delta ^-(\ii\om_n,\vk)=0$ for all $n\geq 0$ and $\vk$, i.e., the Eliashberg equations do not have a non-trivial odd-f gap solution.
This proves No-go theorem~I stated in the main text (note that, there, it only was proved for the Eliashberg equations in the local approximation). Note that this no-go theorem is very general: it is true for {\em any} kind of odd-f SC, i.e., the spin-structure and angular dependence of the gap do not matter.

\subsection{A.2 Spin dependent interactions} 
We now consider a model for a non-conventional superconductor where also interactions mediated by spin-fluctuations are present. In this case, 
the pairing interaction $V$ depends on whether the gap is spin-singlet or spin-triplet, 
\begin{equation} 
\label{Vnonconv1}
V=\begin{cases} V^{\mathrm{s}} = \VC+\Vep-\Vsf & \text{(spin singlet)}\\  V^{\mathrm{t}}=\VC+\Vep+\frac13\Vsf & \text{(spin triplett)} \end{cases} , 
\end{equation} 
with $\Vsf$ the spin-fluctuation interaction, whereas the interaction affecting the normal state is the same in both cases, 
\begin{equation} 
\label{Vnonconv2}
\tilde{V}=\VC+\Vep+\Vsf; 
\end{equation} 
see \cite[Appendix C]{bekaert2018}. 
The spin-fluctuation interaction is given by an expression as in Eq.~\eqref{Vep} but with $\alpha^2F(\Omega;\vk,\vk')$ replaced by another non-negative function proportional to the imaginary part of the spin susceptibility; the only property we need is that $(\Vsf)^-(\ii\om_n,\ii\om_m;\vk,\vk')$ is non-positive, similarly to \eqref{Vep1}.
In the spin-triplet case we have $V^-=(\Vep)^-+\frac13(\Vsf)^-$ and $(\tilde{V}-V)^-=\frac23(\Vsf)^-$, which both are non-positive, and thus \eqref{criterion3} leads to a contradiction, as above. 
However, in the spin-singlet case, $V^-=(\Vep)^- -(\Vsf)^-$ and $(\tilde{V}-V)^-=2(\Vsf)^-$: since $-(\Vsf)^-$ is non-negative, $V^-$ can be positive and thus, in this case, \eqref{criterion3} cannot rule out possible odd-f SC. 

We summarize this result as follows. 

\medskip

\noindent {\bf No-go theorem II.} {\em In $P^*T^*$-invariant Eliashberg theory, spin-triplet odd-f SC is impossible in a one-band superconductor with non-dynamically screened Coulomb interactions and attractive interactions mediated by phonons and spin fluctuations.} 

\section{Appendix B: Solutions of odd-f SC models}
We give details and numeric results demonstrating odd-f SC solutions of the Eliashberg equations \eqref{E1} with interaction potentials 
\begin{equation} 
\label{V2} 
V(\ii\om_n,\ii\om_m) = \lambda f(|\om_n-\om_m|/\Omega) 
\end{equation} 
depending  some function $f(x)$ of $x\geq 0$, together with two parameters $\lambda>0$ and $\Omega>0$. 

\begin{figure}
\includegraphics[width=0.45\linewidth]{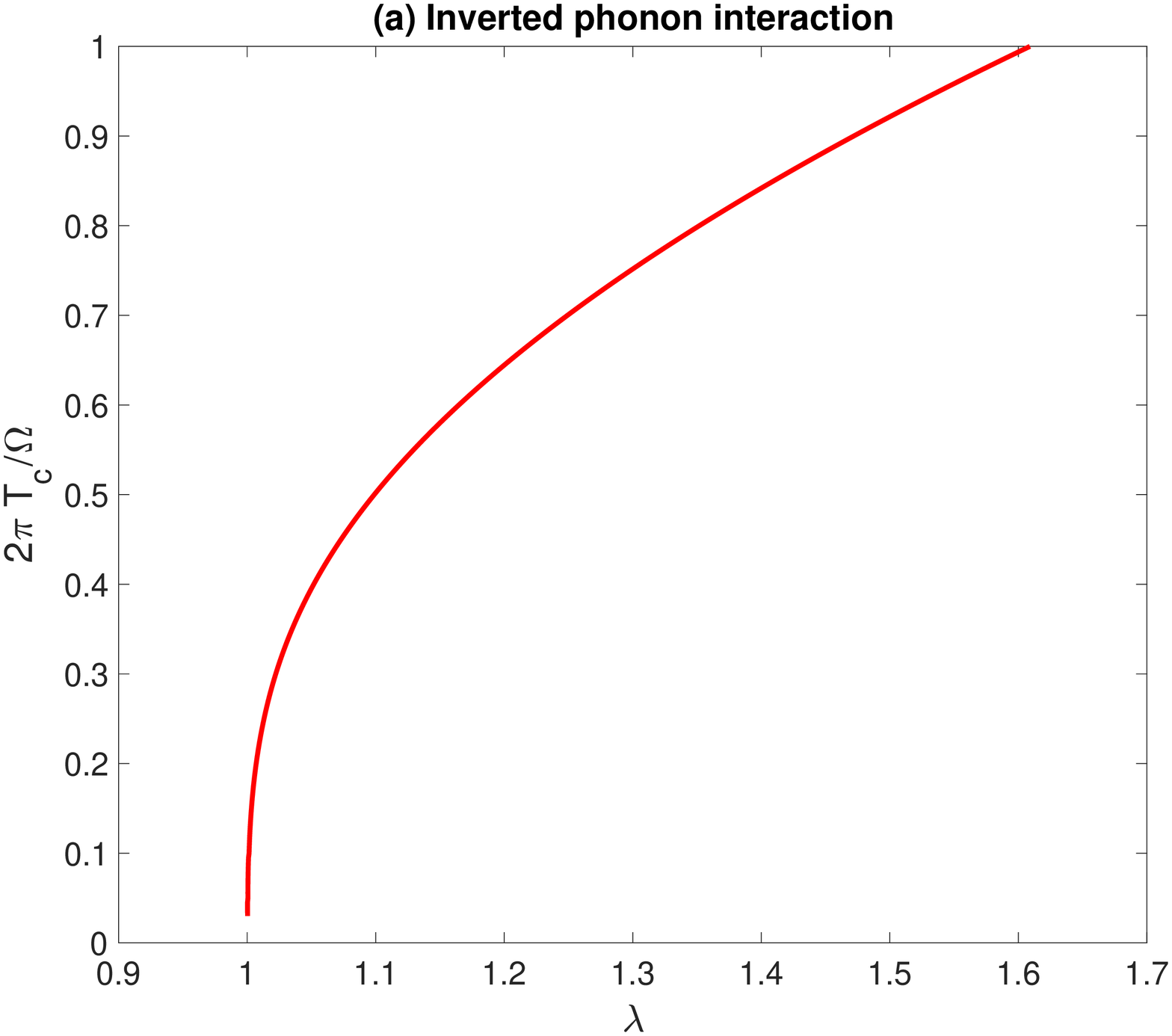} 
\includegraphics[width=0.45\linewidth]{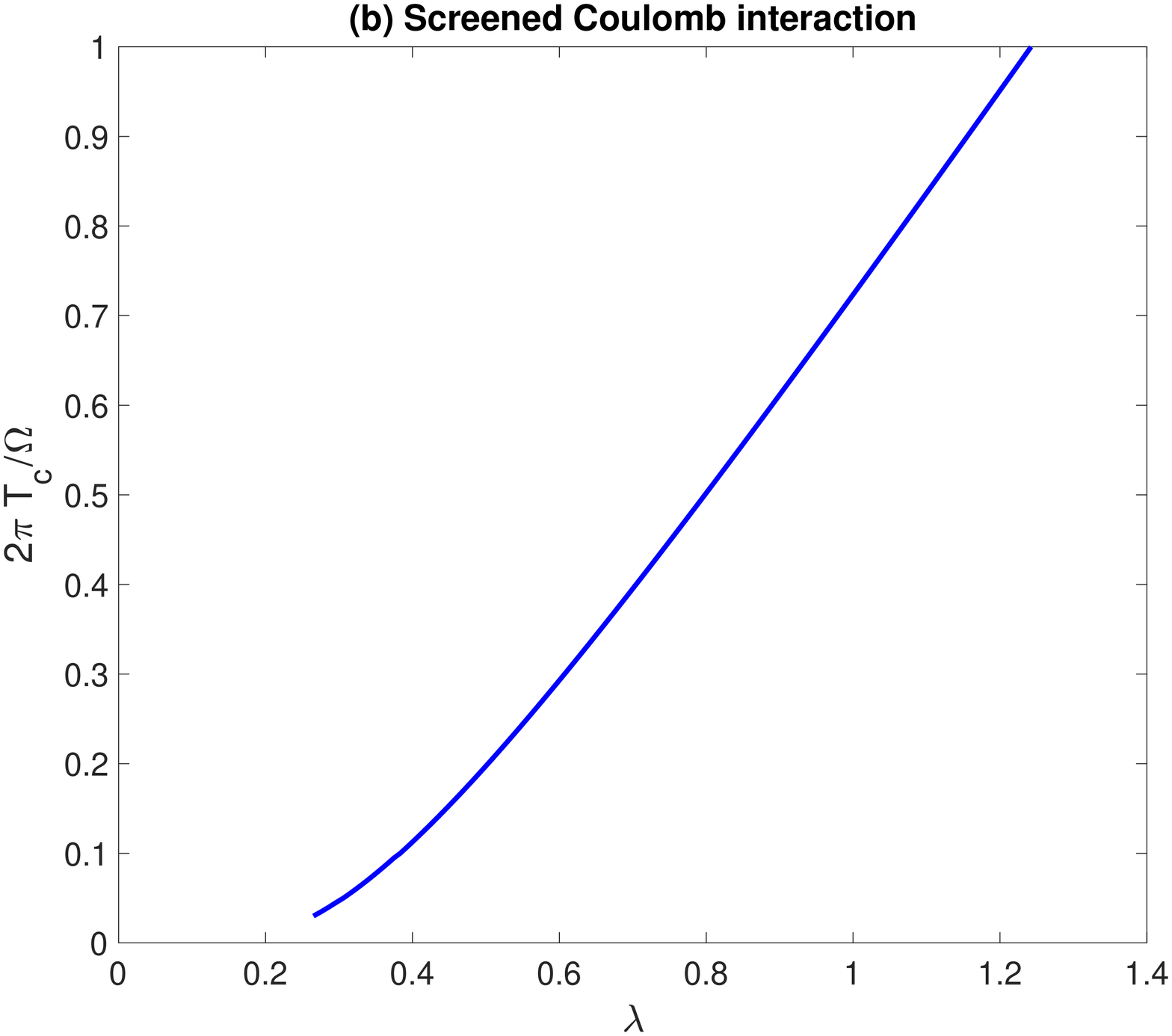}
\caption{Superconducting critical temperature for odd-f SC, $T_c$, versus coupling, $\lambda$, for the Eliashberg equations \eqref{E1} with the interaction potential in \eqref{V2} for two different examples: (a) $f(x)=1/(1+x^2)$ (inverted phonon interaction), (b) $f(x)=\log(1+1/x)$ (screened Coulomb interaction). We checked that, in both cases, $T_c$ grows towards $+\infty$ for $\lambda\to\infty$. The curves end short before $T_c=0$ due to convergence problems of our numeric procedure for $T_c\to 0$.} 
\label{fig1}
\end{figure}

\medskip 

\noindent {\bf Models.} Our first example is $V(\ii\om_n,\ii\om_m)=\lambda \Omega^2/[\Omega^2+(\om_n-\om_m)^2]$,  
which is like a phonon potential with one optical phonon mode  but with the wrong sign 
(this is obtained from \eqref{V0} by setting $\mu^*=0$,  $\alpha^2 F(\Omega)=\lambda(\Omega_0/2)\delta(\Omega-\Omega_0)$ (Dirac delta), 
renaming $\Omega_0\to \Omega$ after the integration, and changing $\lambda$ to $-\lambda$); 
this corresponds to \eqref{V2} with $f(x)=1/(1+x^2)$, and we refer to it as {\em inverted phonon interaction}.  
Our second example is $f(x)=\log(1+1/x)$ which, as we show below following \cite{wu2022}, describes a dynamically screened Coulomb potential close to a quantum critical point; 
we refer to  this as {\em screened Coulomb interaction}. Our third example is $f(x)=1/x^\gamma$ inspired by \cite{abanov2020,wu2022}, as discussed in the main text; we refer to it as {\em inverted $\gamma$-model}. 

To motivate our second example, we start with the potential in \eqref{VQCP} in three dimensions (3D), restrict the momenta to the Fermi surface: $|\vk|=|\vk'|=k_F$ (Fermi momentum) so that $(\vk-\vk')^2=2k_F^2-2k_F^2\cos(\theta)$ with $\theta$ the angle between $\vk$ and $\vk'$, and average over $\theta$: 
\begin{equation}
V(\ii\om_n,\ii\om_m) = N(0)\frac12 \int_0^\pi \frac{g_{\text{eff}}}{[2k_F^2-2k_F^2\cos(\theta)]+\Gamma\frac{|\om_n-\om_m|}{[2k_F^2-2k_F^2\cos(\theta)]^{1/2}}} \sin(\theta)d\theta =
\frac{g_{\text{eff}}N(0)}{6k_F^2}\log\left(1+\frac{8k_F^3}{\Gamma|\om_n-\om_m|} \right)   
\end{equation} 
where, in the second step, we computed an exact integral (the reason for including the DOS is explained in \cite{allen1983}, for example). This proves a claim after \eqref{VQCP} in the main text. Clearly this has the form of a potential in \eqref{V2} with $f(x)=\log(1+1/x)$, $\lambda=g_{\text{eff}}N(0)/6k_F^2$, and $\Omega=8k_F^3/\Gamma$. One can make a similar computation in 2D considered in \cite{wu2022} but, in this case, the resulting function $f(x)$ does not have such a simple explicit formula; thus, for simplicity, we restricted our numeric investigations to the 3D case. 

\medskip 

\noindent {\bf Method.} For the potential $V$ in \eqref{V2}, 
\begin{equation} 
\label{V1minus}
V^-(\ii\om_n,\ii\om_m) = \lambda f(|\om_n-\om_m|/\Omega)-\lambda f(|\om_n+\om_m|/\Omega) . 
\end{equation}  
One can easily check that, if the function $f(x)$ of $x\geq 0$ is monotonically decreasing, then $V^-$ is non-negative;  thus, in such a case, odd-f SC cannot be ruled out by \eqref{criterion1}. 

To show that such a model indeed describes odd-f SC, we compute $T_c$ using \eqref{Tc}. For that, we write \eqref{Tc} in the following way, 
\begin{equation} 
v_n = \lambda\sum_{m\geq 0} M_{n,m}(v_n-v_m),\quad M_{n,m}\equiv \frac{\pi T}{\lambda\om_n} V^-(\ii\om_n,\ii\om_m),\quad v_n\equiv \frac{\Delta^-(\ii\om_n)}{\om_n}; 
\end{equation}  
we find it convenient to define the matrix $M=(M_{n,m})_{n,m\geq 0}$ so that it is independent of the coupling parameter $\lambda$. Inserting \eqref{V1minus}, we find 
\begin{equation} 
\label{Mnm} 
M_{n,m} = \frac1{2n+1}\left[ f(|n-m|X)  -f(|n+m+1|X)\right], \quad X \equiv \frac{2\pi T}{\Omega}, 
\end{equation} 
i.e., $M$ depends only on one parameter $X$. Thus, the $T_c$-equation \eqref{Tc} can be written as eigenvalue equation for the matrix $A=(A_{n,m})_{n,m\geq 0}$ with the matrix elements 
\begin{equation} 
\label{AfromM} 
A_{n,m}\coloneqq \delta_{n,m}\sum_{k\geq 0} M_{n,k}-M_{n,m}, 
\end{equation} 
i.e., 
\begin{equation} 
\sum_{m\geq 0} A_{n,m}v_m = \frac1{\lambda} v_n. 
\end{equation} 

Denoting the largest eigenvalue of the matrix $A$ as $g(X)$, we get the following odd-f $T_c$-equation, 
\begin{equation} 
\label{Tc1}
\frac1{\lambda} = g(2\pi T_c/\Omega).
\end{equation} 
We recall that, for even-f SC, $T_c$ is given by the McMillan formula \cite{mcmillan1968} with has the form $T_c=\Omega\exp(-1/\lambda)$ with $\lambda=V(\ii\om_0,\ii\om_0)$ and $\Omega$ determined by the function $V(\ii\om_n,\ii\om_m)$ in \eqref{V0}; this has exactly the form \eqref{Tc1} with $g(X)=-\log(X)$. Thus, our odd-f $T_c$-equation is similar to a conventional one, except that the function $-\log(X)$ is replaced by the function $g(X)$ obtained by diagonalizing a $X$-dependent infinite matrix $A$. 

To compute $g(X)$, we approximate $M$ by a $N\times N$-matrix, $M_N=(M_{n,m})_{n,m=0}^{N-1}$, compute the corresponding finite matrix $A_N$ from $M_N$ using \eqref{AfromM} with $M$ replaced by $M_N$, compute the larges eigenvalue, $g(X)$, of $A_N$ numerically, and check convergence by plotting $g(X)$ for different values of $N$. Thus, we made sure that all plots in Figs.~\ref{fig1} are numerically accurate. 

It is worth noting that the solution method described above can be used for any potential of the form \eqref{V2}; we did not make any assumption on the sign of $f(x)$.  

\medskip

\noindent {\bf Analytic result for inverted $\gamma$-model.} The inverted $\gamma$-model  corresponds to \eqref{V2} with $f(x)=1/x^{\gamma}$. In this case, the potential in \eqref{V2} is invariant under the scaling transformations $(\lambda,\Omega)\to (s^{-\gamma}\lambda,s\Omega)$ for arbitrary $s>0$. This and our $T_c$-equation \eqref{Tc1} imply $g(X/s)/g(X)=s^{\gamma}$ for all $s>0$ and $X=2\pi T_c/\Omega>0$; the latter equation has a unique solution depending on one constant $C\geq 0$: $g(X)=CX^{-\gamma}$ for all $X>0$. Inserting this in \eqref{Tc1} we obtain 
\begin{equation} 
\label{Tcgamma} 
T_c= K\Omega\lambda^{1/\gamma}\quad (\lambda>0) 
\end{equation} 
with $K=C^{1/\gamma}/2\pi$. 

\medskip 

\noindent {\bf Numeric results.} We computed the function $g(X)$ numerically for various examples. We found that the convergence of our numeric procedure improves with increasing values of $X$. 
In the limit $X\to 0$, we encountered convergence problems, i.e., for $X<0.005$ or so the required values of $N$ to reach convergence were too large for our limited computation resources.  
A related numeric issue for our second and third examples,  $f(x)=\log(1+1/x)$ and $1/x^\gamma$, is that $f(x)$ diverges as $x\to 0$; to solve this problem, we replaced $1/x$ by $1/\sqrt{x^2+\varepsilon^2}$ and varied $\varepsilon$ to make sure that $\varepsilon$ is small enough that results do not change within the desired accuracy if $\varepsilon$ is decreased further. We found that changes in the parameter $\varepsilon$ mainly affect the function $g(X)$ in the regime $X\to 0^+$. Due to these numeric limitations of our method, we were not able to compute $g(X)$ reliably in the limit $X\to 0^+$ using our numeric method. 
However, for the inverted $\gamma$-model, our analytic result shows that $g(0)=+\infty$, which implies that odd-f SC is possible for arbitrarily small $\lambda>0$ for this model. 

We also checked the $T_c$-equation \eqref{Tcgamma} for the inverted $\gamma$-model numerically and found perfect agreement for some value of $K$ we computed numerically; for example, we obtained $K=0.1822$, $0.1911$ and $0.1829$ for $\gamma=0.5$, $1$ and $2$, respectively (numerical values we give are accurate until and including the last digit).  

We also found that, for $\lambda>10$ or so, the numerically computed $T_c$-curves in Fig.~\ref{fig1} can be well approximated by $T_c\approx 0.18 \, \Omega \lambda^{1/2}$ (inverted phonon interaction) and $T_c\approx 0.19\, \Omega \lambda$ (screened Coulomb interaction), respectively; note that, in the former and latter cases, $f(x)\approx 1/x^2$ and $f(x)\approx 1/x$ for large $x$, respectively. 
\medskip 

\noindent {\bf Discussion.} We find it interesting that the $T_c$-curves for odd-f SC and repulsive interactions are qualitatively similar to $T_c$-curves for conventional even-f SC: $T_c$ grows monotonically with the coupling constant $\lambda$. However, for conventional even-f SC, the $T_c$-curve is determined by the universal function $g(X)=-\log(X)$, while for odd-f SC this function $g(X)$ is model dependent. 
For the inverted Coulomb potential, our numerical results suggest that odd-f SC is only possible for $\lambda>\lambda_c$ with $\lambda_c\approx 1>0$ (see Fig.~\ref{fig1}(a)), but for the the inverted $\gamma$-model, $\lambda_c=0$; the latter and our numerics suggest to us that $\lambda_c=0$ also for the screened Coulomb potential (see Fig.~\ref{fig1}(b)). We believe that $\lambda_c>0$ for all potentials $V(\ii\om_n,\ii\om_m)$ which remain finite as $|\om_n-\om_m|\to 0$. 

Our numeric results suggest that the exact $T_c$-formula \eqref{Tcgamma} for the inverted $\gamma$-model captures the leading large-$\lambda$ behaviour of many models: 
if $V(\ii\om_n,\ii\om_m)\approx \lambda(\Omega/|\om_n-\om_m|)^\gamma$ for $|\om_n-\om_m|\to \infty$, then $T_c\approx K\Omega \lambda^{1/\gamma}$ for $\lambda\to\infty$.
\end{widetext} 


\begin{thebibliography}{29}
\expandafter\ifx\csname natexlab\endcsname\relax\def\natexlab#1{#1}\fi
\expandafter\ifx\csname bibnamefont\endcsname\relax
  \def\bibnamefont#1{#1}\fi
\expandafter\ifx\csname bibfnamefont\endcsname\relax
  \def\bibfnamefont#1{#1}\fi
\expandafter\ifx\csname citenamefont\endcsname\relax
  \def\citenamefont#1{#1}\fi
\expandafter\ifx\csname url\endcsname\relax
  \def\url#1{\texttt{#1}}\fi
\expandafter\ifx\csname urlprefix\endcsname\relax\def\urlprefix{URL }\fi
\providecommand{\bibinfo}[2]{#2}
\providecommand{\eprint}[2][]{\url{#2}}

\bibitem[{\citenamefont{Leggett}(2006)}]{leggett2006}
\bibinfo{author}{\bibfnamefont{A.~J.} \bibnamefont{Leggett}},
  \emph{\bibinfo{title}{Quantum liquids: Bose condensation and Cooper pairing
  in condensed-matter systems}} (\bibinfo{publisher}{Oxford university press},
  \bibinfo{year}{2006}).

\bibitem[{\citenamefont{Svistunov et~al.}(2015)\citenamefont{Svistunov, Babaev,
  and Prokof'ev}}]{svistunov2015}
\bibinfo{author}{\bibfnamefont{B.~V.} \bibnamefont{Svistunov}},
  \bibinfo{author}{\bibfnamefont{E.~S.} \bibnamefont{Babaev}},
  \bibnamefont{and} \bibinfo{author}{\bibfnamefont{N.~V.}
  \bibnamefont{Prokof'ev}}, \emph{\bibinfo{title}{Superfluid states of matter}}
  (\bibinfo{publisher}{Crc Press}, \bibinfo{year}{2015}).

\bibitem[{\citenamefont{Arute et~al.}(2019)}]{arute2019}
\bibinfo{author}{\bibfnamefont{F.}~\bibnamefont{Arute}} \bibnamefont{et~al.},
  \bibinfo{journal}{Nature} \textbf{\bibinfo{volume}{574}},
  \bibinfo{pages}{505} (\bibinfo{year}{2019}).

\bibitem[{\citenamefont{Tanaka et~al.}(2011)\citenamefont{Tanaka, Sato, and
  Nagaosa}}]{tanaka2011}
\bibinfo{author}{\bibfnamefont{Y.}~\bibnamefont{Tanaka}},
  \bibinfo{author}{\bibfnamefont{M.}~\bibnamefont{Sato}}, \bibnamefont{and}
  \bibinfo{author}{\bibfnamefont{N.}~\bibnamefont{Nagaosa}},
  \bibinfo{journal}{J. Phys. Soc. Japan} \textbf{\bibinfo{volume}{81}},
  \bibinfo{pages}{011013} (\bibinfo{year}{2011}).

\bibitem[{\citenamefont{Linder and Balatsky}(2019)}]{linder2019}
\bibinfo{author}{\bibfnamefont{J.}~\bibnamefont{Linder}} \bibnamefont{and}
  \bibinfo{author}{\bibfnamefont{A.~V.} \bibnamefont{Balatsky}},
  \bibinfo{journal}{Rev. Mod. Phys.} \textbf{\bibinfo{volume}{91}},
  \bibinfo{pages}{045005} (\bibinfo{year}{2019}).

\bibitem[{\citenamefont{Berezinskii}(1974)}]{berezinskii1974}
\bibinfo{author}{\bibfnamefont{V.~L.} \bibnamefont{Berezinskii}},
  \bibinfo{journal}{JETP Lett} \textbf{\bibinfo{volume}{20}},
  \bibinfo{pages}{287} (\bibinfo{year}{1974}).

\bibitem[{\citenamefont{Balatsky and Abrahams}(1992)}]{balatsky1992}
\bibinfo{author}{\bibfnamefont{A.}~\bibnamefont{Balatsky}} \bibnamefont{and}
  \bibinfo{author}{\bibfnamefont{E.}~\bibnamefont{Abrahams}},
  \bibinfo{journal}{Phys. Rev. B} \textbf{\bibinfo{volume}{45}},
  \bibinfo{pages}{13125} (\bibinfo{year}{1992}).

\bibitem[{\citenamefont{Abrahams et~al.}(1993)\citenamefont{Abrahams, Balatsky,
  Schrieffer, and Allen}}]{abrahams1993}
\bibinfo{author}{\bibfnamefont{E.}~\bibnamefont{Abrahams}},
  \bibinfo{author}{\bibfnamefont{A.}~\bibnamefont{Balatsky}},
  \bibinfo{author}{\bibfnamefont{J.~R.} \bibnamefont{Schrieffer}},
  \bibnamefont{and} \bibinfo{author}{\bibfnamefont{P.~B.} \bibnamefont{Allen}},
  \bibinfo{journal}{Phys. Rev. B} \textbf{\bibinfo{volume}{47}},
  \bibinfo{pages}{513} (\bibinfo{year}{1993}).

\bibitem[{\citenamefont{Pimenov and Chubukov}(2022)}]{pimenov2022}
\bibinfo{author}{\bibfnamefont{D.}~\bibnamefont{Pimenov}} \bibnamefont{and}
  \bibinfo{author}{\bibfnamefont{A.~V.} \bibnamefont{Chubukov}},
  \emph{\bibinfo{title}{Odd-frequency pairing and time-reversal symmetry
  breaking for repulsive interactions}} (\bibinfo{year}{2022}),
  \bibinfo{note}{arXiv:2206.01783 [cond-mat.supr-con]}.

\bibitem[{\citenamefont{Allen and Mitrovi{\'c}}(1983)}]{allen1983}
\bibinfo{author}{\bibfnamefont{P.~B.} \bibnamefont{Allen}} \bibnamefont{and}
  \bibinfo{author}{\bibfnamefont{B.}~\bibnamefont{Mitrovi{\'c}}}, in
  \emph{\bibinfo{booktitle}{Sol. stat. phys.}} (\bibinfo{publisher}{Elsevier},
  \bibinfo{year}{1983}), vol.~\bibinfo{volume}{37}, pp. \bibinfo{pages}{1--92}.

\bibitem[{com()}]{comment}
\bibinfo{note}{For mathematicians we mention that we slightly abuse notation
  here and denote the Fourier transform of the fermion field operators
  $c_{\sigma}({\bf r},t)$ mentioned in the introduction as
  $c_{\sigma}(\ii\om_n,\vk)$, to agree with common notation in the condensed
  matter literature.}

\bibitem[{\citenamefont{Carbotte}(1990)}]{carbotte1990}
\bibinfo{author}{\bibfnamefont{J.~P.} \bibnamefont{Carbotte}},
  \bibinfo{journal}{Rev. Mod. Phys.} \textbf{\bibinfo{volume}{62}},
  \bibinfo{pages}{1027} (\bibinfo{year}{1990}).

\bibitem[{\citenamefont{Marsiglio and Carbotte}(2008)}]{marsiglio2008}
\bibinfo{author}{\bibfnamefont{F.}~\bibnamefont{Marsiglio}} \bibnamefont{and}
  \bibinfo{author}{\bibfnamefont{J.~P.} \bibnamefont{Carbotte}}, in
  \emph{\bibinfo{booktitle}{Superconductivity}} (\bibinfo{publisher}{Springer},
  \bibinfo{year}{2008}), pp. \bibinfo{pages}{73--162}.

\bibitem[{\citenamefont{Marsiglio}(2020)}]{marsiglio2020}
\bibinfo{author}{\bibfnamefont{F.}~\bibnamefont{Marsiglio}},
  \bibinfo{journal}{Ann. Phys. (N. Y.)} \textbf{\bibinfo{volume}{417}},
  \bibinfo{pages}{168102} (\bibinfo{year}{2020}).

\bibitem[{\citenamefont{Fuseya et~al.}(2003)\citenamefont{Fuseya, Kohno, and
  Miyake}}]{fuseya2003}
\bibinfo{author}{\bibfnamefont{Y.}~\bibnamefont{Fuseya}},
  \bibinfo{author}{\bibfnamefont{H.}~\bibnamefont{Kohno}}, \bibnamefont{and}
  \bibinfo{author}{\bibfnamefont{K.}~\bibnamefont{Miyake}},
  \bibinfo{journal}{J. Phys. Soc. Jpn.} \textbf{\bibinfo{volume}{72}},
  \bibinfo{pages}{2914} (\bibinfo{year}{2003}).

\bibitem[{\citenamefont{Wu et~al.}(2022)\citenamefont{Wu, Zhang, Abanov, and
  Chubukov}}]{wu2022}
\bibinfo{author}{\bibfnamefont{Y.-M.} \bibnamefont{Wu}},
  \bibinfo{author}{\bibfnamefont{S.-S.} \bibnamefont{Zhang}},
  \bibinfo{author}{\bibfnamefont{A.}~\bibnamefont{Abanov}}, \bibnamefont{and}
  \bibinfo{author}{\bibfnamefont{A.~V.} \bibnamefont{Chubukov}},
  \emph{\bibinfo{title}{Odd frequency pairing in a quantum critical metal}}
  (\bibinfo{year}{2022}), \bibinfo{note}{arXiv:2205.10903 [cond-mat.str-el]}.

\bibitem[{\citenamefont{Brando et~al.}(2016)\citenamefont{Brando, Belitz,
  Grosche, and Kirkpatrick}}]{brando2016}
\bibinfo{author}{\bibfnamefont{M.}~\bibnamefont{Brando}},
  \bibinfo{author}{\bibfnamefont{D.}~\bibnamefont{Belitz}},
  \bibinfo{author}{\bibfnamefont{F.~M.} \bibnamefont{Grosche}},
  \bibnamefont{and} \bibinfo{author}{\bibfnamefont{T.~R.}
  \bibnamefont{Kirkpatrick}}, \bibinfo{journal}{Rev. Mod. Phys.}
  \textbf{\bibinfo{volume}{88}}, \bibinfo{pages}{039901}
  (\bibinfo{year}{2016}).

\bibitem[{\citenamefont{Abanov and Chubukov}(2020)}]{abanov2020}
\bibinfo{author}{\bibfnamefont{A.}~\bibnamefont{Abanov}} \bibnamefont{and}
  \bibinfo{author}{\bibfnamefont{A.~V.} \bibnamefont{Chubukov}},
  \bibinfo{journal}{Phys. Rev. B} \textbf{\bibinfo{volume}{102}},
  \bibinfo{pages}{024524} (\bibinfo{year}{2020}).

\bibitem[{\citenamefont{Komendov{\'a} et~al.}(2015)\citenamefont{Komendov{\'a},
  Balatsky, and Black-Schaffer}}]{komendova2015}
\bibinfo{author}{\bibfnamefont{L.}~\bibnamefont{Komendov{\'a}}},
  \bibinfo{author}{\bibfnamefont{A.~V.} \bibnamefont{Balatsky}},
  \bibnamefont{and} \bibinfo{author}{\bibfnamefont{A.~M.}
  \bibnamefont{Black-Schaffer}}, \bibinfo{journal}{Phys. Rev. B}
  \textbf{\bibinfo{volume}{92}}, \bibinfo{pages}{094517}
  (\bibinfo{year}{2015}).

\bibitem[{\citenamefont{Asano and Sasaki}(2015)}]{asano2015}
\bibinfo{author}{\bibfnamefont{Y.}~\bibnamefont{Asano}} \bibnamefont{and}
  \bibinfo{author}{\bibfnamefont{A.}~\bibnamefont{Sasaki}},
  \bibinfo{journal}{Phys. Rev. B} \textbf{\bibinfo{volume}{92}},
  \bibinfo{pages}{224508} (\bibinfo{year}{2015}).

\bibitem[{\citenamefont{Komendov{\'a} and
  Black-Schaffer}(2017)}]{komendova2017}
\bibinfo{author}{\bibfnamefont{L.}~\bibnamefont{Komendov{\'a}}}
  \bibnamefont{and} \bibinfo{author}{\bibfnamefont{A.~M.}
  \bibnamefont{Black-Schaffer}}, \bibinfo{journal}{Phys. Rev. Lett.}
  \textbf{\bibinfo{volume}{119}}, \bibinfo{pages}{087001}
  (\bibinfo{year}{2017}).

\bibitem[{\citenamefont{Sukhachov et~al.}(2019)\citenamefont{Sukhachov,
  Juri{\v{c}}i{\'c}, and Balatsky}}]{sukhachov2019}
\bibinfo{author}{\bibfnamefont{P.~O.} \bibnamefont{Sukhachov}},
  \bibinfo{author}{\bibfnamefont{V.}~\bibnamefont{Juri{\v{c}}i{\'c}}},
  \bibnamefont{and} \bibinfo{author}{\bibfnamefont{A.~V.}
  \bibnamefont{Balatsky}}, \bibinfo{journal}{Phys. Rev. B}
  \textbf{\bibinfo{volume}{100}}, \bibinfo{pages}{180502}
  (\bibinfo{year}{2019}).

\bibitem[{\citenamefont{Triola et~al.}(2020)\citenamefont{Triola, Cayao, and
  Black-Schaffer}}]{triola2020}
\bibinfo{author}{\bibfnamefont{C.}~\bibnamefont{Triola}},
  \bibinfo{author}{\bibfnamefont{J.}~\bibnamefont{Cayao}}, \bibnamefont{and}
  \bibinfo{author}{\bibfnamefont{A.~M.} \bibnamefont{Black-Schaffer}},
  \bibinfo{journal}{Ann. Phys. (Berl.)} \textbf{\bibinfo{volume}{532}},
  \bibinfo{pages}{1900298} (\bibinfo{year}{2020}).

\bibitem[{\citenamefont{Hainzl and Seiringer}(2008)}]{hainzl2008}
\bibinfo{author}{\bibfnamefont{C.}~\bibnamefont{Hainzl}} \bibnamefont{and}
  \bibinfo{author}{\bibfnamefont{R.}~\bibnamefont{Seiringer}},
  \bibinfo{journal}{Physical Review B} \textbf{\bibinfo{volume}{77}},
  \bibinfo{pages}{184517} (\bibinfo{year}{2008}).

\bibitem[{\citenamefont{Hainzl and Seiringer}(2016)}]{hainzl2016}
\bibinfo{author}{\bibfnamefont{C.}~\bibnamefont{Hainzl}} \bibnamefont{and}
  \bibinfo{author}{\bibfnamefont{R.}~\bibnamefont{Seiringer}},
  \bibinfo{journal}{J. Math. Phys.} \textbf{\bibinfo{volume}{57}},
  \bibinfo{pages}{021101} (\bibinfo{year}{2016}).

\bibitem[{\citenamefont{Langmann et~al.}(2019)\citenamefont{Langmann, Triola,
  and Balatsky}}]{triola2019}
\bibinfo{author}{\bibfnamefont{E.}~\bibnamefont{Langmann}},
  \bibinfo{author}{\bibfnamefont{C.}~\bibnamefont{Triola}}, \bibnamefont{and}
  \bibinfo{author}{\bibfnamefont{A.~V.} \bibnamefont{Balatsky}},
  \bibinfo{journal}{Phys. Rev. Lett.} \textbf{\bibinfo{volume}{122}},
  \bibinfo{pages}{157001} (\bibinfo{year}{2019}).

\bibitem[{\citenamefont{Salmhofer}(2007)}]{salmhofer2007}
\bibinfo{author}{\bibfnamefont{M.}~\bibnamefont{Salmhofer}},
  \emph{\bibinfo{title}{Renormalization: an introduction}}
  (\bibinfo{publisher}{Springer Science \& Business Media},
  \bibinfo{year}{2007}).

\bibitem[{\citenamefont{Bekaert et~al.}(2018)\citenamefont{Bekaert, Aperis,
  Partoens, Oppeneer, and Milo{\v{s}}evi{\'c}}}]{bekaert2018}
\bibinfo{author}{\bibfnamefont{J.}~\bibnamefont{Bekaert}},
  \bibinfo{author}{\bibfnamefont{A.}~\bibnamefont{Aperis}},
  \bibinfo{author}{\bibfnamefont{B.}~\bibnamefont{Partoens}},
  \bibinfo{author}{\bibfnamefont{P.~M.} \bibnamefont{Oppeneer}},
  \bibnamefont{and} \bibinfo{author}{\bibfnamefont{M.~V.}
  \bibnamefont{Milo{\v{s}}evi{\'c}}}, \bibinfo{journal}{Phys. Rev. B}
  \textbf{\bibinfo{volume}{97}}, \bibinfo{pages}{014503}
  (\bibinfo{year}{2018}).

\bibitem[{\citenamefont{McMillan}(1968)}]{mcmillan1968}
\bibinfo{author}{\bibfnamefont{W.}~\bibnamefont{McMillan}},
  \bibinfo{journal}{Phys. Rev.} \textbf{\bibinfo{volume}{167}},
  \bibinfo{pages}{331} (\bibinfo{year}{1968}).

\end{thebibliography}

\end{document}